# Seventy Years of the EPR Paradox

Marian Kupczynski

*Department of Mathematics and Statistics, University of Ottawa, 585,av. King-Edward,Ottawa.Ont.K1N 6N5*
*and*
*Département de l'Informatique, UQO, Case postale 1250 ,succursale Hull, Gatineau. Quebec, Canada J8X 3X 7*

**Abstract**. In spite of the fact that statistical predictions of quantum theory (QT) can only be tested if large amount of data is available a claim has been made that QT provides the most complete description of an individual physical system. Einstein's opposition to this claim and the paradox he presented in the article written together with Podolsky and Rosen in 1935 inspired generations of physicists in their quest for better understanding of QT. Seventy years after EPR article it is clear that without deep understanding of the character and limitations of QT one may not hope to find a meaningful unified theory of all physical interactions, manipulate qubits or construct a quantum computer.. In this paper we present shortly the EPR paper, the discussion, which followed it and Bell inequalities (BI). To avoid various paradoxes we advocate purely statistical contextual interpretation (PSC) of QT. According to PSC a state vector is not an attribute of a single electron, photon, trapped ion or quantum dot. A value of an observable assigned to a physical system has only a meaning in a context of a particular physical experiment PSC does not provide any mental space-time picture of sub phenomena. The EPR paradox is avoided because the reduction of the state vector in the measurement process is a passage from a description of the whole ensemble of the experimental results to a particular sub-ensemble of these results. We show that the violation of BI is neither a proof of the completeness of QT nor of its non-locality. Therefore we rephrase the EPR question and ask whether QT is "predictably "complete or in other words does it provide the complete description of experimental data. To test the "predictable completeness" it is not necessary to perform additional experiments it is sufficient to analyze more in detail the existing experimental data by using various non-parametric purity tests and other specific statistical tools invented to study the fine structure the time-series.

# INTRODUCTION

. It is not an easy task to open a session on the Nature of the Quantum World since as Bohr said there is no quantum world but only an abstract quantum mechanical description.[1]. If we do not accept Bohr's point of view we leave the science and we start telling the stories about twin photons communicating instantaneously or about two fair and random dices giving always correct matching outcomes. In this talk we want to show that all such stories are not necessarily true and that they are only examples of imprecise language, incorrect mental images and analogies used nowadays by some physicists.

To explain properties of visible objects and phenomena we introduce some more elementary invisible objects (elementary particles, atoms, molecules, electromagnetic waves etc) and some invisible sub phenomena. In classical physics visible and invisible objects were characterized by some attributes which could be discovered and measured by some perfect instruments and not modified by the act of the measurement. Mental models of invisible sub phenomena were like miniature versions of phenomena in the visible world giving us a comfort of intuitive understanding.

When in some experiments a beam produced by some source behaved once as a beam of particles having well defined individuality and once as a material wave and we were unable to reconcile these two apparently contradictory pictures of invisible sub phenomena the Quantum Theory (QT) was created [1]. The theory provided only abstract algorithms explaining in a quantitative probabilistic way various physical phenomena. Any attempt to complete this abstract formalism by some familiar and universal mental picture of invisible sub phenomena failed so far thus we lost the intuitive understanding of the world.

Let us list various types of physical experiments:

1) SOURCE -------BEAM--------APPARATUS--------NEW SEPARATED BEAMS
2) SOURCE -------BEAM--------APPARATUS--------COUNTERS-----TIME- SERIES OF COUNTS
3) BEAM 1+BEAM 2---------INTERACTION--------NEW BEAMS
4) BEAM 1+BEAM 2---------INTERACTION--------COUNTERS – TIME -SERIES OF COUNTS
5) LASER-----PULSE------SYSTEM-------NEW PULSE---- COUNTERS----- TIME- SERIES

In the experiments 1-4 some sources are producing several copies of the "same" physical system. In the experiment 1 the initial beam after passing by the apparatus is split into several new beams, which can be used as new sources for further experimentation. In the experiment 2 the only outcome is a time- series of counts from various detectors placed behind the apparatus. In the experiment 3 various copies of two physical systems are put into interaction (for example collisions of some elementary particles), new systems emerge after this interaction and can be used for further experimentation. The experiment 4 is the same as the experiment 3 but the only outcome is a time- series of counts from some detectors used to analyze the outgoing beams. The experiment 5 schematizes recent experiments with trapped ions, electrons etc. A physical system in a trap is prepared for the experiment by specific initial manipulations. A short pulse from a laser is projected on the system and a modified outgoing pulse is observed, analyzed and some time- series of final results is recorded. Next the initial conditions in a trap are reset and the experiment is repeated.

In all these experiments no single result is predictable however in a long time series of counts statistical reproducible regularities are observed. QT is used to explain these regularities.

QT gives statistical predictions for distributions of the results obtained in long runs of one experiment or in several repetitions of the same experiment on a single physical system. It is unclear and not obvious how and in what sense a claim can be made that QT provides a complete description of individual physical systems. Einstein never agreed with this claim in spite of the fact that the proof of incompleteness of QT, he gave in the famous EPR paper [2] written together with Podolsky and Rosen, was rejected by Bohr. EPR failed to prove that QT was incomplete but they showed clearly that the assumption that a wave function is an attribute of a single physical system leads to a contradiction called the EPR paradox. Let us review quickly EPR reasoning.

## EPR PARADOX

Let us start with a list of some axioms of quantum mechanics, which are believed to be true, or which at least were believed to be true before the publication of the EPR paper [2]

S1: Any state of an individual physical system is described by a specific wave function $\Psi$
S2: A specific self-adjoint operator $\hat{A}$ represents corresponding physical variable A
S3: Any result of the measurement of A is one of the eigenvalues of $\hat{A}$.
S4: Any measurement of A causes the system to jump into one of the eigenstates of $\hat{A}$.
S5: If a system is in a state described by a wave function $\Psi$ which is not an eigenstate of $\hat{A}$ then a result of any single measurement performed on the system cannot be predicted and only the probabilities of obtaining particular experimental results can be calculated..
S6: There is no state of a physical system in which exact values of two real dynamical observables represented by non-commuting operators can be known simultaneously in particular if we knew exact position of a system its linear momentum would be completely unknown and vice versa.
S7: A wave function $\Psi$ provides a complete description of a pure state of an individual physical system.

EPR considered two particular individual systems I+II in a pure quantum state, which interacted in the past, separated and evolved freely afterwards. On a basis of S4 EPR found that:
1) If a single measurement of an observable A was performed on one of the systems (for example on the system I), and the result was recorded, then the wave function assigned to the pure state of the system II was $\Phi_A$.

2) If a different complementary observable B was chosen to be measured on the system I instead of A, then from any outcome of the measurement of the observable B we would find that the wave function assigned to the same pure state of the system II should be $\Phi_B$ which would be necessarily different than any wave function $\Phi_A$ found in the point 1) since A and B were represented by non-commuting operators.

Thus it was possible to assign two different wave functions $\Phi_A$ and $\Phi_B$ to the same physical reality (the second system after the interaction with the first) without disturbing an any way the system II what together with S1 contradicted S7. This contradiction is real. In statistical interpretation of QT the assumptions S1 and S4 are modified and the EPR paradox avoided. .

For a moment we continue to review more specific arguments given by EPR. Consider two particles I+II such that for t=T: 1) p1+p2 = p and x2 – x1 =q 2) If we measure p1 for the particle I we get immediately p2=p- p1 3) If we measure x1 for the particle I we get immediately : x2= x1+q . Thus by measuring the linear momentum or the position of the particle I without in any way disturbing the particle II we can know the linear momentum or the position of the particle II with an arbitrary precision what contradicts S6. Therefore EPR concluded that the quantum mechanical description of the physical reality was incomplete.

Bohr [3] promptly reacted to EPR paper. Bohr noticed how difficult was to obtain the information about the position or thelinear momentum of the second particle from the knowledge of the result of one measurement performed on the first particle. In particular he considered a single couple of point-like particles, which collided with a two-slit diaphragm, and a single measurement performed at the end on one of these particles. To be able to get information about the linear momentum of the second particle from the measurement of the linear momentum of the first particle he had to use a moving diaphragm and to measure its linear momentum before and after the collision with this particular couple of particles. In order to be able to get information about the position of the second particle from the measurement of the position of the first particle the diaphragm had to be kept fixed and any knowledge of the total linear momentum was lost. Therefore Bohr refuted EPR conclusion that it was possible to assign two different wave functions to the same reality (the second system after the interaction with the first) since the different functions could be assigned only in different experiments in which <u>both</u> systems were exposed to different influences. Bohr's arguments showed that different eigenfunction expansions of the same wave function $\Psi$ described a behavior of the same physical system in different mutually excluding (complementary) experiments. Therefore $\Psi$ was not an attribute of a single couple of these systems but only a mathematical tool used to deduct statistical regularities of various experimental data. Bohr clearly demonstrated that one might not measure simultaneously the position and the linear momentum of a particle with a precision violating Heisenberg uncertainty relations but by no means he proved that the QT provided complete description of the individual quantum system. The very use of the mental picture (model) of a point- like particle having a definite linear momentum contained in its definition a hidden variable its position in some moment of time. Of course this position was unknown but a point-like particle had to be somewhere. Clearly Bohr noticed the inconsistency of this picture of sub phenomena therefore in his subsequent essays [1] he insisted that in any experimental arrangement suited for the studying of proper quantum phenomena we had to do not with an ignorance of the value of certain physical quantities but with impossibility of defining these quantities in unambiguous way and that any quantum phenomenon should be considered as a whole.

In 1936 Furry [4] made important comment which underlined the contextual character of QT: "The measuring instruments must always be included as part of the physical situation from which our experience is obtained … a system is not any independent seat of real attributes simply because it has ceased to interact dynamically with other systems" . Schrödinger [5] gave a detailed analysis of EPR states and his statement: "No matter how far apart the particles are when we try to collect one of them the relative probabilities of finding it in different places are strongly affected by the "interference term" in the cross-section it is not really "free".. " is considered nowadays as a first precise definition of the entanglement. We see immediately that to explain the correlated behavior of the entangled pair we have three options. The first is to assume that all the couples are the same and that the random behaviors of the members of each couple during the measurements are correlated by some spooky action at a distance. The second is to assume that we have a statistical ensemble of different couples and for each couple the results of all measurements are predetermined and correlated in the moment of production. The third is to assume that the long-range correlations of the results of the measurements performed on the beams are due to their particular preparation by a common source but at the same time assume that the result of any single measurement is not predetermined. Einstein opted for the second option and Bohr chose the extreme variant of the third option claiming that the quantum phenomenon should be described as a whole and that there is no consistent picture of sub phenomena possible.

Let us analyze the implications of the second option If particles carry some correlated information which predetermines the outcomes of the distant measurements then each couple is described by some correlated supplementary parameters called hidden variables and the description of phenomena given by QT is incomplete. Therefore Einstein concluded that the essentially statistical character of contemporary QT was solely ascribed to the fact that (this theory) operated with an incomplete description of physical systems. For Einstein, Fock, Landau, Blokhintzev and many other physicists the wave function did not describe the individual physical system but only a whole ensemble of identically prepared physical systems and their probabilistic behavior observed in the experiments was due to the lack of the control and the knowledge of some supplemental parameters describing invisible deterministic sub phenomena. This purely statistical interpretation (SI) of QT, reviewed and used extensively by Ballentine [6,7] , makes plausible the existence of hidden variable models of sub phenomena capable to reproduce the predictions of QT. Unfortunately we may mention here only few further contributions to the fascinating discussion on the foundations of QT started by EPR.

In 1952 Bohm [8] presented a spin version of the EPR paradox (EPR-B) and showed that the pilot-wave model proposed first time by de Broglie [9] was able to reproduce all non-relativistic predictions of QT. The pilot -wave description of the EPR experiment being non-relativistic was also non-local.

Trying to find a local description of the EPR-B Bell [10,11] studied so called local realistic hidden variable models (LRHV) and arrived to the important conclusion that all LRHV led to the inequalities called now Bell inequalities (BI) and that these inequalities had to be violated by the predictions of QM for some spin polarization correlation experiments (SPCE). In 1969 Clauser, Horne, Shimony and Holt [12] gave so called CHSH version of BI and showed explicitly how CHSH could be tested in SPCE.

In 1970 Wigner[13]made an important observation that BI do apply not only to EPR type experiments but to any sequence of spin polarization measurements performed on one beam. It was the first indication that the non-contextual character of LRHV was more important in various proofs of BI than the locality assumption.

In 1974 Clauser and Horne [14] gave a new very general proof of BI and in 1978 Clauser and Shimony gave excellent review [15] of the experimental evidence of the violation of BI existing at that time. The subsequent accurate experiments, in particular those by Aspect et al. [16,17] and more recently by Weihs et al. [18] , seemed to close the remaining loop-holes and confirmed clearly that the predictions of QT for SPCE were confirmed and Bell and CHSH inequalities violated. Therefore it was proven that the description given by QT of SPCE might not be completed by any LRHV what by some physicists is being considered even now or as a proof of the completeness of QT or as a proof that QT violates the locality. of physical interactions. Not everybody agreed with this conclusion.

In 1986 Aerts[19] showed that non-Kolmogorovian character of the quantum probabilities pointed out in 1981 by Accardi [20] was due to the indeterminacy on the measurements in contrast to the indeterminacy on the initial states assumed in LRHV and gave the examples of classical correlation experiments violating BI. In 1982 Pitovsky [21,22] constructed a local non –Kolmogorovian hidden variable model which after slight reformulation [23,24] became explicitly contextual and allowed to reproduce all quantum mechanical polarization predictions for SPCE. In 1984 we showed [23,25] that various proofs of BI are questionable and the violation of BI did imply neither the completeness of QT nor its non-locality and we proposed [23,26] completely new tests of completeness of QT Seeing that our papers [23-26] were unknown or not understood we refined and extended our arguments in recent papers [27-29], which we review in subsequent sections. Several authors [30-34] came recently to the similar conclusions.

## STATISTICAL INTERPRETATION AND EPR-B

Let us list here main assumptions of SI [7,8], which after adding the assumption 5 we call PSC.
- 1.A state vector $\Psi$ is not an attribute of a single electron, photon, trapped ion, quantum dot etc. A state vector $\Psi$ or a density matrix $\rho$ describe only an ensemble of identical state preparations of some physical systems
- 2.A state vector $\Psi$ together with a hermitian or self-adjoint operator $\hat{O}$ representing an observable O provide the probability distribution of the outcomes of the measurement of the observable O obtained for the sequence of preparations described by $\Psi$. For simplicity of presentation we assume here that $\hat{O}$ has only a discrete spectrum.
- 3. If we want to perform subsequent experiments but only on those physical systems from the original ensemble for which a nondestructive measurement of the observable O gave the same result $o_i$ then this whole new sub ensemble of systems is described by a reduced state vector $\Psi_i$ which is an eigenfunction of the self -adjoint operator $\hat{O}$ representing the observable O such that $\hat{O}\Psi_i = o_i\Psi_i$. In SI a mysterious wave function reduction is neither instantaneous nor non- local .

- 4.For EPR experiment a state vector describing the system II obtained by the reduction of the entangled state of two physical system I+II describes only the sub-ensemble of the systems II being the partners of those systems I for which the measurement of some observable gave the same specific outcome. To different outcomes correspond different sub- ensembles described by different reduced state vectors therefore there is no EPR paradox.

- 5.A value of a physical observable associated with a pure quantum ensemble and in this way with an individual physical system being its member, is not an attribute of the system revealed by a measuring apparatus; but a characteristic of this ensemble created by its interaction with the measuring device [24,28,34]

We see that any experiment is described by a couple ($\Psi$, $\hat{O}$) or ($\rho$, $\hat{O}$). The QT is a contextual theory since it predicts only probability distributions. The probabilities are neither attributes of individual physical systems nor attributes of measuring devices but only characteristics of a whole random experiment.[24,27,28]. The QT is not a theory of quantum individual systems but it is a theory of quantum phenomena.

Let us examine schematically EPR-B realized in SPCE [16-18].

A pulse from a laser hitting a non-linear crystal produces two correlated physical fields propagating with constant velocities in opposite directions. Each of these fields has a property that it produces clicks when hitting the photon detector far away. We place two polarization analyzers A and B in front of the detectors on both sides and after interaction of the fields with the analyzers we obtain two correlated time- series of clicks on the far away detectors. Each analyzer is characterized by its macroscopic direction vectors, which may be changed at any time. By changing the direction vectors we have various coincidence experiments labeled by (**A**, **B**) where **A** and **B** are the direction vectors for the analyzers A and B respectively.

In QT the crystal is described as a source of couples of photons in a total spin zero state described by a state vector.

$$\Psi = (|+\rangle_P |-\rangle_P - |-\rangle_P |+\rangle_P)/\sqrt{2}. \qquad (1)$$

where $|+\rangle_P$ and $|-\rangle_P$ are state vectors corresponding to photon states in which their spin is "up" or "down" in the direction **P** respectively**.**

The EPR-B paradox may be presented as follows [8]. If we measure a spin projection of any photon I on the direction **P** we have an equal probability to obtain a result + or −. In any case the two photon state vector $\Psi$ is reduced "instantaneously". If we obtain + the reduced state vector of the photon II is $|-\rangle_P$, if we obtain - the reduced state vector of the photon II is $|+\rangle_P$. By choosing a direction **P** for the measurement to be performed on the photon I, when "the photons are in flight and far apart" we can assign different incompatible reduced state vectors to the same photon II. In other words: we can predict with certainty, and without in any way disturbing the second photon, that the **P**-component of the spin of the photon II must have the opposite value to the value of the measured **P**- component of the spin of the photon I. Therefore for any direction **P** the **P**-component of the spin of the photon II has unknown but predetermined value what contradicts QT. The EPR-B can be refuted by two different arguments.

- The first argument [29] is similar to that given by Bohr [3]. By saying that the photons had a time to separate we assume a mental image of two point-like photons, which are produced and which after some time become separated and free. We do not see any particular couple of these photons and we do not follow its space-time evolution. We record only the clicks on the far away coincidence counters. To be able to deduce the value of a particular spin projection for the photon II from the measurement made on the photon I we should have had for each experiment a different experimental design (impossible to realize) giving us much more information about each individual couple of the photons than we have in a simple coincidence experiment.

- The second argument is based on SPC. The reduced one particle state $|+\rangle_P$ describes only the whole ensemble of the partners of the particles I which were found to have "spin down" by an analyzer P pointing the direction **P.** For various directions **P** it is a different sub ensemble of particles II. Correlations are due to the common history not because of the non-locality of interactions [24,27].According to SPC a state $\Psi$ allows only to predict statistical predictions for the correlations observed in a long run of various experiments with different couples (A, B) of spin polarization analyzers characterized by their macroscopic direction vectors (**A**, **B**) with respect to some standard reference frame. Since the sharp macroscopic directions do not exist we may only represent the analyzers **A** and **B** by some probability distributions of microscopic directions $d\rho_A(a)$ on $O_A = \{a | |1-a\bullet A|<\varepsilon_A\}$ and $d\rho_B(b)$ on $O_B = \{b | |1-b\bullet B|<\varepsilon_B\}$. Therefore the probability P(x,y|**A**,**B**) where x=±1 and y=±1 denote "spin up" or "spin down" in the directions **A** and **B** respectively (each value of x and y corresponds to clicks on corresponding counters) is given by:

$$P(x,y|\mathbf{A},\mathbf{B})= \eta(\mathbf{A})\,\eta(\mathbf{B}) \int\int p(x,y|\mathbf{a},\mathbf{b})\,d\rho_{\mathbf{A}}(\mathbf{a})\,d\rho_{\mathbf{B}}(\mathbf{b}) \qquad (2)$$

where $p(x,y|\mathbf{a},\mathbf{b})$ are calculated using QT, the integration is done over $O_{\mathbf{A}} \times O_{\mathbf{B}}$ and $\eta(\mathbf{A})$ and $\eta(\mathbf{B})$ are some counting efficiency factors. In spite of the fact that $p(x,x|\mathbf{a},\mathbf{a})=0$ for any direction of $\mathbf{a}$ the probabilities $P(x,x|\mathbf{A},\mathbf{A})$ are never equal to 0 even if the counters are perfect. Therefore contrary to a general belief the QT does not predict strict anti-correlations of the time-series of clicks in any experiment (A,A) [24,27,29]

In spite of the fact that the long-range correlations are not perfect they do exist and as Bell said they cry for explanation. One way to explain them is to complete SI by a LRHV model of the sub phenomena.

Different arguments in favor of SI may be found in [7,8,30-34]

## LHRV MODEL AND BELL INEQUALITIES

According to LHRV a source is producing EPR pairs. In the moment of production a state of a pair is completely determined by a hidden parameter $\lambda \in \Lambda$ and both members of each pair have unknown but well defined and strictly correlated spin projection values in all directions, distributed according to some unknown joint probability distribution $\rho(\lambda)$. The measuring devices act independently and locally and each of these devices can register a correct value or fail to register it with a small probability [14]. The probability distributions for all different coincidence experiments (A,B) are obtained by "conditionalization" from a single sample (probability) space $\Lambda$. In other words each experiment is described in a standard way by its own marginal probability distribution obtained from $\rho(\lambda)$ [25,27,29].

In LHRV the macroscopic directions of analyzers (**A**, **B**) are assumed to be sharp (the efficiency factors $\eta(\mathbf{A})= \eta(\mathbf{B})=1$) or absorbed in the probability functions $p_1(x|\mathbf{A},\lambda)$ and $p_2(y|\mathbf{B},\lambda)$ and the formula (2) is rewritten in the following form [11,14]

$$P(x,y|\mathbf{A},\mathbf{B})= \int_\Lambda p_1(x|\mathbf{A},\lambda)p_2(y|\mathbf{B},\lambda)\,\rho(\lambda)\,d\lambda \qquad (3)$$

The probabilistic model described by the formula (3) leads to BI-CHSH inequalities, which are violated by the predictions of QT and by the results of several experiments [16-18]. Since formula (3) seemed to be very general and based only on the locality assumption it was concluded that the violation of BI [16-18] proved that QT was non-local.

Careful analysis [27,29] shows that the formula (3) is very restrictive. The probability distribution of any random experiment (**A**, **B**) can be always reproduced by using another formula:

$$P(x,y|\mathbf{A},\mathbf{B})= \int_{\Lambda(A,B)} p(x,y|\mathbf{A},\mathbf{B},\lambda)\,\rho_{AB}(\lambda)\,d\lambda \qquad (4)$$

The passage from the formula (4) to the formula (3) is only possible if $\Lambda(A,B)=\Lambda$, $\rho_{AB}(\lambda) = \rho(\lambda)$ and if for any value of $\lambda$ and for all directions

$$p(x,y|\mathbf{A},\mathbf{B},\lambda)= p_1(x|\mathbf{A},\lambda)p_2(y|\mathbf{B},\lambda) \qquad (5)$$

The assumptions: $\Lambda(A,B)=\Lambda$, $\rho_{AB}(\lambda) = \rho(\lambda)$ were incorrectly believed to be implied by the locality of the measurements. They violate contextuality of QT and have far reaching implications namely the assumption (5) and the formula (3) are justified only in two situations:

1) In the first the $p_1(x,y|\mathbf{A},\lambda)$ and $p_2(x,y|\mathbf{B},\lambda)$ are equal 0 or 1 [10] and for each value of $\lambda$ there is a strict determinism and we have a statistical ensemble of the classical particles for which in any moment of time the values of the spin polarization projection in any direction are predetermined or

2) In the second for any value of $\lambda$ the random variables giving the outcomes of x and y are independent for all directions **A** and **B** and we have a statistical mixture with respect to $\lambda$ of completely uncorrelated measurements which destroy totally the correlations which might have existed between the couples before these measurements.

For any theoretical probabilistic model there is a random experiment with a well-defined protocol, which is described by this model [25,27,29]. Let us examine the situation 1 and specify a random experiment described by the probabilistic model given by the formula (3). The protocol contains 3 steps.

i) A spin state of a couple is completely described by its variable $\lambda$. A mixed statistical ensemble E is created by taking $n_1$ pairs described $\lambda_1$, $n_2$ pairs described by $\lambda_2$ and $n_k$ pairs described by $\lambda_k$.

ii) A pair is drawn at random from the ensemble E and the spin polarization projection is measured on each member of the couple in all available directions (**A**, **B**), the results are recorded and a couple is replaced into E.

iii) Using the data from a long run of this random experiment the probabilities P(x,y|**A**,**B**) for each particular pair of directions are estimated from the empirical joint probability distribution.

It is obvious that different SPCE experiments are mutually exclusive and cannot be replaced by a single random experiment described above. The formula (2) given by QT, and the Pitovsky model [21-24] of sub phenomena able to reproduce it, are both contextual in contrast to LRHV. Leaving aside a problem of the existence and the utility of deterministic local and contextual hidden variable models we discuss now the completeness of QT.

## PREDICTABLE COMPLETENESS AND PURITY TESTS.

As was mentioned in introduction any physical experiment performed on an ensemble of identically prepared systems or repeated several times on the same "individual' physical system in a trap gives at the end several time series of numerical data corresponding to so called different " runs" of the experiment. Let us represent this in a schematic way:

$$\text{System (S)} \longrightarrow \text{Experiment (E)} \longrightarrow T(S,E,i) \qquad (7)$$

where $T(S,E,i)$ is a time- series obtained in i-th run of an experiment E performed on a system or systems S.

A central notion in QT is a system prepared in a pure state described by its state vector $\Psi$. In SPC an individual system is in a pure state only if it is a member of some pure ensemble of identically prepared systems. The operational definition of pure ensemble was given in [35]

A pure ensemble is an ensemble characterized by such empirical distributions of various counting rates, which remain approximately unchanged for any rich sub ensembles drawn from this ensemble in a random way.

In any hidden variable model of sub phenomena a pure quantum ensemble becomes a mixed statistical ensemble with respect to some uncontrollable variables.

Any sub ensemble of a pure ensemble is indistinguishable from the initial ensemble. The sub ensembles of the mixed ensemble can differ between them. Since we do not control the distribution of hidden variables the time-series $T(S,E,i)$ may differ from run to run of the same experiment. Therefore the non parametric statistical purity tests on $T(S,E,i)$ provide tests of the completeness of QT [23,26,27]. Several years ago, in the different context, we introduced and studied several purity tests, which might be used for this purpose [36-38].

The "predictable completeness" means a bit more. QT gives predictions for probability distributions of outcomes of various experiments performed on physical systems. The probability distribution averages out the information contained in $T(S,E,i)$. If in some experiments we detect reproducible fine structure in $T(S,E,i)$ which was not and could not be predicted by QT we will prove that QT is not predictably complete.

## CONCLUSIONS

By emphasizing the difference between quantum phenomena and hypothetical invisible sub phenomena we showed that paradoxes are only found if incorrect models of sub phenomena are used. The violation of BI –CHSH demonstrated clearly that "an entangled pair of photons" resembled neither "a pair of Bertlmann's socks" (look pp.139-158 in [11]) nor "a pair of fair and random dices". The description of the sub phenomena must be contextual any individual experimental outcome cannot be predetermined by a value of some variable assigned only to a physical system but it depends also on the variables describing the devices used in this particular experiment [24,34].

As we saw the contextual character of QT was strongly underlined by the fathers of QT but its completeness was taken for granted and after seventy years the EPR question about the completeness of QT is still unanswered. We don't even know whether statistical description SPC given by QT of quantum phenomena is predictably complete. There are already perhaps some indicators that it is not a case. For example it was demonstrated by Claude Cohen-Tannoudji and collaborators [39] that to describe effectively the behavior of cold trapped ions the continuous quantum evolution had to be supplemented by some quantum jumps obeying some stochastic Lévy process. It is also well known that to be able to explain experimental data QT has to be complemented by some unproven semi–empirical approximations and models containing free parameters. This flexibility of QT is quite worrying [35] because the theory becomes non falsifiable. The difficulties to reconcile QT with the theory of gravitation are perhaps beneficial because they force us to reexamine the epistemological foundations and implications of both theories what will lead probably to a significant progress in understanding of the phenomena in the surrounding us world.

We would like to repeat a statement from [35] "Let us be more critical of the models we propose, of the conclusions we obtain, and let us check the operational status of the language we use to deal with data".